# Approximate mass spectra of the heavy mesons under a Coulomb plus logarithmic spin-dependent potential function.

E. Omugbe[1], E. P Inyang[2], K. Adegoke[1] and E. M. Khokha[3]

[1]Department of Physics and Engineering Physics, Faculty of Science, Obafemi Awolowo University Ile-Ife, Osun State, Nigeria.
[2]Department of Physics, National Open University of Nigeria, Abuja, Nigeria.
[3]Faculty of Computer Science and Engineering, King Salman International University (KSIU), South Sinai, El Tur, Egypt.

[1]Corresponding author email: eomugbe@oauife.edu.ng

## Abstract

In this paper, we presented an approximate analytical treatment of the Coulomb plus logarithmic potential using time-independent perturbation theory to investigate the mass spectra of bottomonium and charmonium mesons for the low-order quantum states. Within the framework of the Schrödinger equation, we derived the energy equation to first-order corrections and employed it to model the free potential parameters through fitting to experimental data of the Particle Data Group. The calculated bottomonium masses exhibited good agreement with experimental values, yielding an absolute percentage average deviation (APAD) of 0.25%, which improves upon previously reported theoretical results where the Pekeris type approximation was used to deform the centrifugal barrier and potential. Similarly, the vector and pseudoscalar charmonium masses were obtained with an APAD of 1.69%, demonstrating improved and comparable accuracy relative to existing competing theoretical calculations Our results of the energy spectra to first-order corrections in the absence of spin components gave small percent errors in comparison with exact numerical solutions obtained using the matrix Numerov method.

## 1. Introduction

The theoretical study of charmonium and bottomonium mass spectra has been a central topic in hadronic physics, with non-relativistic potential models serving as the dominant framework. The Cornell quark–antiquark potential is extensively used to describe meson bound states. The potential in the presence of centrifugal barrier and spin-dependent functions does not admit exact analytical solution and must be solved approximately or numerically. The bound state calculations of the Cornell potential in momentum-space have been obtained using a simplified subtraction method, combined with Lagrange polynomial interpolation and the Nyström method [1]. The resulting energy eigenvalues where found to be accurate. Kang and Won [2] applied the Crank–Nicolson method with imaginary time evolution to obtain accurate numerical solutions of the Schrödinger equation under the Cornell potential achieving accurate eigenvalue of up to eleven significant digits. Also, in a recent contribution, Christas and Dhir [3] investigated the spectroscopic properties of quarkonium families using a Cornell-type potential with perturbative spin-dependent terms. They solved the Schrödinger equation numerically to obtained masses, Regge trajectories, and decay widths. A similar approach was used by Mistry et al. [4], who investigated, the heavy-flavor mesons of the type $Q\bar{Q}$ ($Q = c$, $b$) and obtained the Regge trajectories in the ($J$, $M^2$) and ($n$, $M^2$) planes; including Quantum Chromo dynamics (QCD) correction factors. They reported leptonic, photonic, and gluonic decay widths for a range of quantum states. Sreelakshmi and Ranjan [5] employed a non-relativistic framework, combining a Coulomb-like confinement term with a one-gluon-exchange screening term perturbed by spin-dependent corrections to calculate the charmonium mass spectra and decay properties, benchmarking their results against both experimental data and competing theoretical predictions. Garg et al. [6] carried out a detailed analysis of bottomonium mass spectra and decay properties by incorporating spin interactions into a non-relativistic quark-antiquark potential and solving the Schrödinger equation via the Runge–Kutta method. The decay widths for di-leptonic, di-gamma, tri-gamma, and di-gluon channels were obtained with the Van Royen–Weisskopf formula, and radiative transition widths were computed to shed light on the non-perturbative structure of QCD. Maireche [7] investigated the bound-state solutions of diatomic molecules and the mass spectra of heavy quarkonium using an improved modified Kratzer potential combined with a screened Coulomb potential within a three-dimensional non-commutative phase-space framework. By incorporating non-commutative phase-space symmetries, the model revealed corrections to the energy eigenvalues and mass spectra, demonstrating that non-commutative effects play a significant role in describing the quantum behavior of molecular and heavy quark systems. In a recent work, Bokade and Bhaghyesh [8] treated the bottomonium meson within the relativistic framework using a screened potential, where they reported the mass spectra for $s$, $p$, $d$ and $g$ waves together with E1 and M1 transition widths and the leptonic decay widths of $s-d$ mixed states. Sultan et al. [9], compared the relativized Godfrey–Isgur quark model with an unquenched regime and found that the softening of the quark-antiquark wavefunction improves the description of bottomonium decay. The Nikiforov–Uvarov (NU) method has been widely adopted as an analytical technique for obtaining the exact or approximate solutions of the Schrödinger equation, and several groups have applied it to quarkonium systems with different potential choices. The influence of spatial dimensionality was explored in Ref. [10], where the Killingbeck and inversely quadratic potentials were solved via the NU method in N-dimensional space; beyond reproducing experimental mass spectra, the authors analyzed how temperature and dimensionality shape the thermodynamic properties of quarkonium. Abualkishik and Al-Jamel [11] introduced a relativistic approach within the Nikiforov–Uvarov (NU) framework by solving the Klein–Gordon equation in D dimensions, incorporating a scalar–vector quadratic confinement together with a modified screened Yukawa potential. By fitting their model to experimental data, they compared the results obtained from fitting the charmonium and bottomonium mesons independently with those derived from a unified parameter set applied jointly to both mesons. They reported unobserved $1f$ state masses for the heavy mesons. Within the relativistic regime, Badalov et al. [12] solved the Dirac equation with a generalized tanh-shaped potntial and reported charmonium and bottomonium masses for different quantum states. The analytical solution of the Klein–Gordon oscillator was obtained in Ref. [13] in the presence of cosmic string spacetime, rainbow gravity, and external magnetic field. The resulting energy equations for both quarkonium were solved analytically for two choices of rainbow function, and the associated thermodynamic quantities showed markedly different behavior depending on the choice. The analytical solutions of the Klein–Gordon equation for arbitrary $l$-states using a modified effective mass under the modified unequal scalar and vector Coulomb–Hulthén potential has been investigated within the framework of relativistic non-commutative three-dimensional real space. By employing an approximation to the centrifugal term, together with Bopp's shift method and perturbation theory, Maireche [14] derived an analytical expressions for the energy eigenvalues and the mass spectra of charmonium and bottomonium. The predicted results showed good agreement with existing theoretical studies, demonstrating the effectiveness of the proposed model in describing relativistic bound-state systems.

Also, the spinless Salpeter equation solved with a spin-dependent Cornell potential via the Wentzel-Kramers-Brillouin (WKB) approximation, was applied to obtain the mass spectra for charmonium, bottomonium, and bottom-charmed mesons[15]. The relativistic Bethe–Salpeter equation was used to investigate the leptonic decay widths of $s$-wave vector families of $J/\psi$ and $\Upsilon$ heavy-flavor mesons via single-photon annihilation process into lepton pairs. The $1s$-$4s$ mass spectra and leptonic widths were reproduced well for the lowest states, while moderate deviations at higher radial excitations were ascribed to the sensitivity of nodal structures and the omission of higher-order radiative and relativistic corrections [16]. The effect of non-trivial spatial topology and curved geometry on quarkonium spectra has attracted notable interest. In Ref. [17], a conical background with a topological defect was combined with a singularized harmonic potential, and the Schrödinger equation was recast as a bi-confluent Heun equation. The resulting mass spectra and decay constants up to the 4s state agreed with experiment only when the defect parameter was far smaller than unity. In an additional context Omugbe et al. [18] examined the expectation values of the Cornell potential using both direct integration and the Hellmann–Feynman theorem, showing that meson masses and principal quantum numbers shape the mean values and probability densities in ways that are fully consistent with the Heisenberg, Cramér-Rao, and Fisher uncertainty relations. The light-front quark model (LFQM) has been used for connecting quarkonium mass spectra with structural observables. Heavy pseudoscalar and vector mesons were studied in Ref. [19] using a Coulomb-plus-logarithmic potential within the LFQM where the authors identified a newly observed resonance and found that the predicted twist-two distribution amplitudes and branching ratios were in good agreement with lattice QCD simulations and experiment data. Harjapradipta et al. [20] computed the light-front wave functions (LFWFs) by a variational method with harmonic-oscillator basis functions and derived the electromagnetic form factors that compare favorably with other models and recent lattice QCD data. In these contributions to the spectroscopic properties of the heavy mesons, the competing methods and potential functions utilization allows for the improvement of the mesons masses in comparison with observed data of the Particle Data Group (PDG) [21]. Most of the methods utilized numerical approach by either employing QCD inspired potential functions or phenomenological potentials with perturbative spin components [22-28]. A phenomenological potential allows for the empirical modeling of the potential parameters from experimental data points to obtain best fits for the meson properties. However, analytical solutions are few since the most utilized potential functions do not admit exact analytical solution. Several attempts have been made to obtain approximate solution by deforming the centrifugal barrier and the potential function using a Pekeris type approximation [15, 22, 25]. In this present work, we aim to derive an approximate analytical mass spectra equation using the perturbation theory. We proposed the combined Coulomb potential (one-gluon exchange characterized by asymptotic freedom at short distances) and a logarithm function which grows with distance $r$ and confines the quark. The potential function is given

$$V\left(r\right)=-\frac{V_0}{r}+BIn\left(\lambda r\right) \tag{1}$$

where, $V_0$ and $B$ are arbitrary parameters. However, for application to meson mass spectrum, we may substitute a colour factor for the Coulomb parts ( $V_0=4\alpha_s/3$ ) and the parameters must then be obtained from fitting of the approximate mass spectra equation to observed data. In Figure. 1, the potential function for the heavy mesons show both asymptotic freedom via one gluon exchange between the quark and its anti-particle and a logarithm function accounting for confinement phenomena. The remaining parts of this article are organized as follows. In section 2, we will present the approximation solution of the mass spectra via the perturbation method. To confirm the accuracy of the solutions, we will employ matrix Numerov numerical scheme [29]. In section 3, we will present discussion on the heavy mesons mass spectra and benchmark our solutions with the experimental masses of the PDG and other competing theoretical results. In section 4, we will give the conclusion.

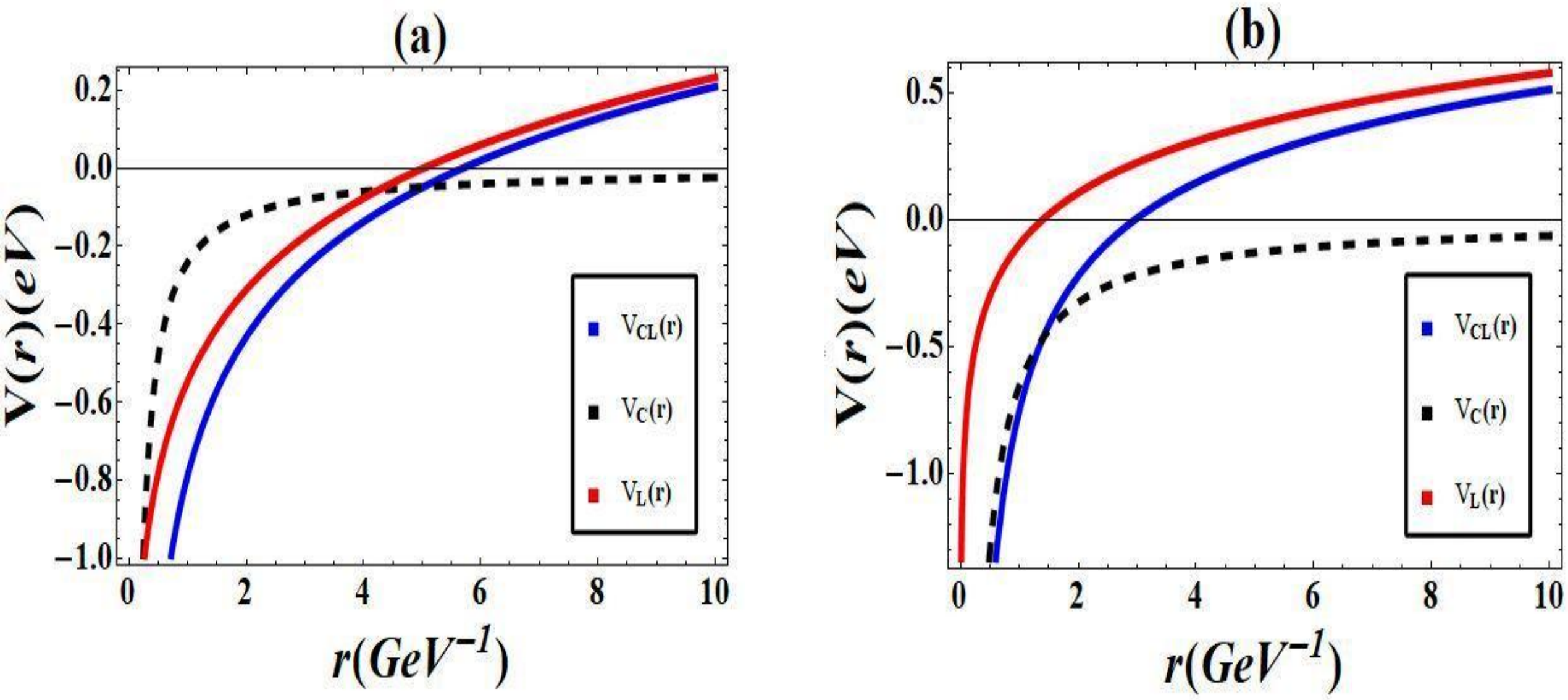

Figure 1.Potential energy curves for (a) Bottomonium meson (b) Charmonium meson. CL=Coulomb +logarithm, C=Coulomb, L= logarithm.

2. **Approximate Solution of the mass spectrum equation via perturbation method**

In this section, we adapt the Coulomb-logarithm potential to the spin-spin, spin-orbit and tensor components following the works in Refs [**15**, **24**, 26, 27, 30, **31**]

$$V_{S-S}(r)=\frac{2}{3m_q m_{\bar{q}}}\nabla^2 V_V(r)\bar{S}\bullet\bar{S}=\frac{32\pi\alpha_s}{9m_q m_{\bar{q}}}\delta^3(r)\bar{S}\bullet\bar{S} \tag{2}$$

$$V_{S-L}(r)=\frac{1}{2m_q m_{\bar{q}}r}\left(3\frac{dV_V(r)}{dr}-\frac{dV_S(r)}{dr}\right)\bar{L}\bullet\bar{S} \tag{3}$$

$$V_T(r)=\frac{4}{m_q m_{\bar{q}}r^3}\bar{T} \tag{4}$$

Choosing $V_V(r)=-4\alpha_s/3r$ as the vector potential and $V_S(r)=B\ln(r)$ as a scalar function, we obtained the modeled potential as

$$V(r)=-\frac{4\alpha_s}{3r}+BIn(\lambda r)+\Lambda_s e^{-\sigma^2 r^2}-\frac{B\langle\bar{\boldsymbol{L}}\cdot\bar{S}\rangle}{2m_q m_{\bar{q}}r^2}+\frac{2\alpha_s\left(2\langle\bar{T}\rangle+\langle\bar{\boldsymbol{L}}\cdot\bar{S}\rangle\right)}{m_q m_{\bar{q}}r^3} \tag{5}$$

The notations $\Lambda_s$, $\langle\bar{\boldsymbol{L}}\cdot\bar{S}\rangle$ and $\langle\bar{T}\rangle$ are the respective spin-spin, spin-orbit and tensor matrix elements given by

$$\Lambda_s=\frac{16\alpha_s\pi}{9m_q m_{\bar{q}}}\left(\frac{\sigma}{\sqrt{\pi}}\right)^3\left(s(s+1)-\frac{3}{2}\right) \tag{6}$$

$$\langle\bar{\boldsymbol{L}}\cdot\bar{S}\rangle=\frac{1}{2}\big(j(j+1)-l(l+1)-s(s+1)\big) \tag{7}$$

$$\langle\bar{T}\rangle=\begin{cases}-\dfrac{l}{6(2l+3)}, & j=l+s\\[2ex] \dfrac{1}{6}, & j=l\\[2ex] -\dfrac{l+1}{6(2l-1)}, & j=l-s\end{cases} \tag{8}$$

The notation $\sigma$ is the Gaussian width arising from the smearing of the delta function in the spin-spin component of the model potential and the $\lambda$ is a scaling factor introduced to the $\ln(\lambda r)$ function to ensure unit consistency in the potential. The quark and anti-quark masses are denoted by $m_q$ and $m_{\bar{q}}$ while $s$ represents the spin quantum number. The quantum number $J^{PC}$ identifies the states with $P((-1)^{l+1})$ and $C((-1)^{l+s})$ as parity and charge conjugation respectively. The $j(l+s)$ is the total angular momentum quantum number. The spin-spin component is responsible for the s-wave ($l=0$) hyperfine splitting between the triplet ($s=1$) and singlet ($s=0$) states. For $l>0$, with $j=l\pm 1, j=l$, the quantum states split into triplets for $p$, $d$, $f$ and $g$ orbitals. The notations $n$ and $l$ represent the principal and orbital momentum quantum numbers respectively. In eq.(**1**), only the Coulomb part of the potential is exactly solvable for any $l$-states, hence we would assume the other terms as perturbing functions and apply the perturbation theory. We would not give full attention to the treatment of this theory but the interested readers may refer to the references [**32**, **33**] for more details.

The radial Schrodinger equation may be constructed as an eigenvalue problem

$$\left(H^{(0)}+\varepsilon H^{'}\right)\left(\psi_{n_r l}^{(0)}(r)+\varepsilon\psi_{n_r l}^{(1)}(r)+\cdots\right)=\left(E_{n_r l}^{(0)}+\varepsilon E_{n_r l}^{(1)}+\cdots\right)\left(\psi_{n_r l}^{(0)}(r)+\varepsilon\psi_{n_r l}^{(1)}(r)+\cdots\right) \tag{9}$$

After expanding eq. (6) and comparing coefficients of powers of $\varepsilon$ to first-order corrections, the unperturbed eigenvalue equation and the first-order corrections to the energy spectra are obtained as

$$H\psi_{n_r l}^{(0)}(r)=E_{n_r l}^{(0)}\psi_{n_r l}^{(0)}(r) \tag{10}$$

$$E_{n_r l}^{(1)}=\int_0^{\infty}\left(\psi_{n_r l}^{(0)}(r)\right)^{*}H^{'}\psi_{n_r l}^{(0)}(r)\,dr \tag{11}$$

The notations $n_r$ and $l$ are the respective radial and orbital angular momentum quantum numbers. To solve the potential in (**5**), we grouped the terms $-\frac{4\alpha_s}{3r}-\frac{B\langle \bar{\boldsymbol{L}}\cdot\bar{S}\rangle}{2m_q m_{\bar{q}} r^2}$ as the reference unperturbed aspect of the Hamiltonian whose energy spectra can be exactly found. This choice would ensure that the spin-orbit interaction is not suppressed in a way that the energy spectra become degenerate within the perturbation approach. The other terms $BIn(\lambda r)+\Lambda_s e^{-\sigma^2 r^2}+\frac{2\alpha_s\left(2\langle\bar{T}\rangle+\langle\bar{\boldsymbol{L}}\cdot\bar{S}\rangle\right)}{m_q m_{\bar{q}} r^3}$ are chosen as the reference perturbed Hamiltonian since there is no exact analytical solution. The bound states solution of the unperturbed equation may be obtained from the series solution method [**34**]

$$E_{n_r l}^{(0)}=-\frac{8\alpha_s^2\mu}{9\hbar^2\left(n_r+\frac{1}{2}+k\right)^2} \tag{12}$$

$$\psi_{n_r,l}^{(0)}(r) = C_{n_r,l} r^{k+\frac{1}{2}} e^{-\omega r} L_{n_r}^{2k}(2\omega r), \tag{13}$$

where

$$k = \sqrt{\left(l+\frac{1}{2}\right)^2 - \frac{\mu B \langle \bar{\boldsymbol{L}} \cdot \bar{S} \rangle}{m_q m_{\bar{q}} \hbar^2}}, \ \omega = \sqrt{-\frac{2\mu E_{n_r l}^{(0)}}{\hbar^2}}$$

The normalization constant ( $C_{n_r l}$ ) in the wave function is obtained from the probability condition:

$$C_{n_r l}^2 \int_0^\infty \left| \psi_{n_r l}^{(0)}(r) \right|^2 dr = 1 \tag{14}$$

$$C_{n_r l} = \sqrt{\frac{(2\omega)^{2k+2} n_r!}{\Gamma(n_r + 2k + 1)(2n_r + 2k + 1)}} \tag{15}$$

Having obtained the unperturbed bound state solutions, the first-order corrections to the energy spectra is obtained using Eq. (**11**)

$$E_{n_r l}^{(1)} = \left\langle \psi_{n_r l}^{(0)*} \left| H^{'} \right| \psi_{n_r l}^{(0)} \right\rangle = \int_0^\infty \left| \psi_{n_r l}^{(0)} \right|^2 H^{'} dr = \sum_{i=1}^{3} I_{n_r,i} \tag{16}$$

where

$$I_{n_r,1} = B \int_0^\infty In(\lambda r) \left| \psi_{n_r l}^{(0)} \right|^2 dr \tag{17}$$

$$I_{n_r,2} = \Lambda_s \int_0^\infty e^{-\sigma^2 r^2} \left| \psi_{n_r l}^{(0)} \right|^2 dr \tag{18}$$

$$I_{n_r,3} = \frac{2\alpha_s \left( 2\langle \bar{T} \rangle + \langle \bar{\boldsymbol{L}} \cdot \bar{S} \rangle \right)}{m_q m_{\bar{q}}} \int_0^\infty \frac{1}{r^3} \left| \psi_{n_r l}^{(0)} \right|^2 dr \tag{19}$$

The integrals in (**17**) -(**19**) are evaluated with the Mathematica program for quantum number $n_r = 0$ and $n_r = 1$;

$$I_{0,1} = B\left( In\left(\frac{\lambda}{2\omega}\right) + \Psi(0, 2+2k) \right), \tag{20}$$

$$I_{0,2}=\frac{\Lambda_s\omega}{\Gamma(2k+2)\sigma^2}\left(\frac{2\omega}{\sigma}\right)^{2k+1}\left(\sigma\Gamma(k+1)\,{}_1F_1\left(k+1,\frac{1}{2},\frac{\omega^2}{\sigma^2}\right)-2\omega\Gamma\left(k+\frac{3}{2}\right){}_1F_1\left(k+\frac{3}{2},\frac{3}{2},\frac{\omega^2}{\sigma^2}\right)\right) \tag{21}$$

$$I_{0,3}=\frac{2\alpha_s\left(2\langle\bar{T}\rangle+\langle\bar{\boldsymbol{L}}\cdot\bar{S}\rangle\right)}{m_q m_{\bar{q}}}\frac{4\omega^3}{4k^3-k} \tag{22}$$

$$I_{1,1}=B\left(\frac{3}{2k+3}+In\left(\frac{\lambda}{2\omega}\right)+\Psi(0,2+2k)\right) \tag{23}$$

$$I_{1,2}=Q_0\left(\sigma\Gamma(1+k)\left(Q_1+4\sigma\omega^2(k+1)Q_2\right)-2\omega\Gamma\left(k+\frac{3}{2}\right)\left(Q_3+Q_4+Q_5\right)\right) \tag{24}$$

where

$$Q_0=\frac{\Lambda_s\omega}{(3+2k)\sigma^4\Gamma(2+2k)}\left(\frac{2\omega}{\sigma}\right)^{2k+1}$$

$$Q_1=(\sigma+2k\sigma)^2\,{}_1F_1\left(k+1,\frac{1}{2},\frac{\omega^2}{\sigma^2}\right)$$

$$Q_2={}_1F_1\left(k+2,\frac{1}{2},\frac{\omega^2}{\sigma^2}\right)+2(2k+1)\,{}_1F_1\left(k+2,\frac{3}{2},\frac{\omega^2}{\sigma^2}\right)$$

$$Q_3=2\sigma^2(2k+1)\,{}_1F_1\left(k+\frac{3}{2},\frac{1}{2},\frac{\omega^2}{\sigma^2}\right)$$

$$Q_4=(\sigma+2k\sigma)^2\,{}_1F_1\left(k+\frac{3}{2},\frac{3}{2},\frac{\omega^2}{\sigma^2}\right)$$

$$Q_5=2\omega^2(3+2k)\,{}_1F_1\left(k+\frac{5}{2},\frac{3}{2},\frac{\omega^2}{\sigma^2}\right)$$

$$I_{1,3}=\frac{8\omega^3\alpha_s\left(2\langle\bar{T}\rangle+\langle\bar{\boldsymbol{L}}\cdot\bar{S}\rangle\right)}{m_q m_{\bar{q}}\left(4k^3-k\right)} \tag{25}$$

where ${}_1F_1(\bullet)$ and $\Psi(\bullet)$ are the respective confluent hypergeometric and polygamma functions

## 3. Discussion of results

The approximate analytical solution of the mass spectra equation to first-order perturbation has been obtained using the equation

$$M_{n_r l} = m_q + m_{\bar{q}} + E_{n_r l}^{(0)} + E_{n_r l}^{(1)} \tag{26}$$

We used the bottom and charm masses based on extensive research works on quark model where the effective masses are given in the ranges $1.2 < m_c < 1.8GeV$ and $4.5 < m_b < 5.4GeV$ [22, 23, 24, 25, 27, 30, 35. 36] and fitted the equation (26) to the masses of the PDG [21]. This procedure resulted into nonlinear equations which we have solved using minimization and the NLPSolve function in Maple program (See eq. $A_2$ and the Maple code in the appendix) to obtain the potential parameters given in Table 1. It is noteworthy to state that eq. ($A_2$) possesses many local minima for fixed fitting points. This implies that there are correlations of parameters which admit nearly the same solution. However, the best fit parameters are obtained based on the lowest APAD values that closely reproduce available experimental masses and other theoretical predictions. We obtained the bottomonium masses given in Table 2 and calculated the absolute percentage average deviation (APAD) with a value of 0.25% from experimental masses given by the relation:

$$APAD\% = \frac{100}{K}\sum_{i}^{K}\left|1-\frac{M_i^{theo.}}{M_i^{\exp.}}\right| \tag{27}$$

where $K$, $M_i^{exp}$ and $M_i^{theo}$ are the numbers of data points, available experimental masses from the particle data group, and calculated masses respectively. The masses obtained with the perturbation method gave improvement over the masses calculated in Ref. [15], [22] and [25] where the Pekeris-type approximation was used to deform the centrifugal barrier. Similarly, we obtained the potential parameters for the charmonium meson given in Table 1. The obtained masses having an absolute percentage average deviation of 1.69% (see Table 3) are in agreement with experimental results and showed improvement over theoretical masses reported in Refs. [15, 24]. In addition, we observed an anomaly in the ordering of the spin-dependent masses in comparison with the PDG values. Our results show the masses ordering of $\boldsymbol{\chi(n^3L_2) < \chi(n^3L_1)} < \chi(n^3L_0)$ which is an inversion of the experimental values. We attribute the ordering to the dominance of the scalar part of the spin-orbit force arising from the logarithmic function confinement. This confinement ( $-B\left\langle \bar{\boldsymbol{L}}\cdot\bar{S}\right\rangle \big/ 2m_q m_{\bar{q}} r^2$ ) causes the energy to be more attractive for higher magnitudes of the total angular momentum quantum number ($j$) and increases the masses for smaller $j$ values. Also, the energy contributions from the $B\ln(\lambda r)$ decrease with increase in $j$, whereas the energy contributions from the spin-spin ( $\Lambda_s e^{-\sigma^2 r^2}$ ) and $2\alpha_s\left(2\left\langle \bar{T}\right\rangle + \left\langle \bar{\boldsymbol{L}}\cdot\bar{S}\right\rangle\right)\big/ m_q m_{\bar{q}} r^3$ functions increase with $j$ but are too weak to push the masses up for higher $j$. The inversion of the masses are also present in higher $l$ states as reported in existing works [26, 36].Generally, our results are consistent with values obtained from other theoretical approaches for higher $l$-states. The absolute deviation of the masses from observed values are given in Table 4 where deviations are smaller for the bottomonium mesons compared to the charmonium spectra. Evidence suggests that the bottomonium meson has lower deviations and APAD values given in the Tables 2 and 4, indicating that the heavier masses of the bottom quarks forming the meson bound state are well described by non-relativistic quantum mechanics compared with the charm quark with a lighter mass. Using arbitrary parameters of the potential, we compared the energy eigenvalues with the ones obtained via the matrix numerov approach and obtain good agreement (see Tables **5**, **6** and **7)**. Generally, the energy spectra increase with the quantum numbers but decrease with the increase in parameters $B$ and $\boldsymbol{\alpha_s}$. It is worth stating that in the absence of spin interactions, the perturbation theory imposes a constraint on the magnitude of the $B$ parameter which must be chosen to be small ($B \ll \boldsymbol{\alpha_s}$). For $B > \boldsymbol{\alpha_s}$, the current perturbation energies would be inaccurate but we may obtain accurate energies for larger $B$ if we assume the logarithmic part of the potential to be the parent component while taking the Coulomb part as the perturbing function. This case becomes difficult to solve analytically for any $l$-state solutions. However, the s-wave energies ($l = 0$) may be obtained with the semi-classical WKB approximation but the wave function is a challenge making numerical solutions such as the matrix numerov approach [**29**] as a viable option. The absolute percentage deviations of the perturbation energies from numerical solutions ($\%\ \text{error} = 100\times\left(\left|1-E^{approx.}/E^{num.}\right|\right)$) are reasonable, further indicating that the first-order corrections to the energy spectra are sufficient and reliable.

**Table 1.** Calculated parameters of interaction potential function.

| | Charmonium $m_c = m_{\bar{c}} = 1.520$ | Bottomonium $m_b = m_{\bar{b}} = 4.890$ |
|---|---|---|
| Fitting points | $\boldsymbol{J/\psi\,(1^3S_1)}$, $\boldsymbol{\eta_c(1^1S_0)}$, $\chi_{c_2}\boldsymbol{(1^3P_2)}$, $\boldsymbol{\psi(2^3S_1)}$, $\chi_{c_1}\boldsymbol{(2^3P_1)}$, $\boldsymbol{h_c(1^1P_1)}$, $\boldsymbol{\eta_c(2^1S_0)}$, $\chi_{c_1}\boldsymbol{(1^3P_1)}$, $\chi_{c_0}\boldsymbol{(1^3P_0)}$ | $\boldsymbol{\Upsilon(1^3S_1)}$, $\boldsymbol{\eta_b(1^1S_0)}$, $\chi_{\boldsymbol{b_0}}\boldsymbol{(2^3P_1)}$, $\chi_{\boldsymbol{b_0}}\boldsymbol{(1^3P_0)}$ |
| $\boldsymbol{\alpha_s}$ | 0.4910 | 0.1811 |
| $\boldsymbol{\lambda(GeV)}$ | 0.7144 | 0.1991 |
| $\boldsymbol{B(GeV)}$ | 0.2932 | 0.3383 |
| $\boldsymbol{\sigma(GeV)}$ | 13.8523 | 5.0638 |

**Table 2**. Approximate mass spectrum of bottomonium meson in $GeV$.

| State | | Present | Ref. [15] | Ref. [28] | Ref. [24] | Ref. [25] | Ref. [22] | Expt. [21] |
|---|---|---|---|---|---|---|---|---|
| $n^{2s+1}L_j$ | $J^{PC}$ | | | | | | | |
| $\Upsilon(1^3S_1)^*$ | $1^{--}$ | 9.41983 | 9.906 | 9.465 | 9.460 | 9.49081 | 9.525 | $9.460 \pm (2.6 \times 10^{-4})$ |
| $\eta_b(1^1S_0)^*$ | $0^{-+}$ | 9.41555 | 9.916 | 9.402 | 9.398 | 9.43601 | 9.472 | $9.399 \pm (2 \times 10^{-3})$ |
| $\Upsilon(2^3S_1)$ | $1^{--}$ | 9.96051 | 10.240 | 10.003 | 10.023 | 10.01257 | 10.049 | $10.023 \pm (3.1 \times 10^{-4})$ |
| $\eta_b(2^1S_0)$ | $0^{-+}$ | 9.95998 | 10.251 | 9.976 | 9.990 | 9.99146 | 10.028 | |
| $\chi_{b_1}(1^3P_1)$ | $1^{++}$ | 9.90778 | 9.906 | 9.876 | 9.892 | 9.87371 | 9.875 | $9.893 \pm (2.6 \times 10^{-4}) \pm (3.1 \times 10^{-4})$ |
| $\chi_{b_2}(1^3P_2)$ | $2^{++}$ | 9.90022 | 9.912 | 9.897 | 9.912 | 9.89083 | 9.903 | $9.912 \pm (2.6 \times 10^{-4}) \pm (3.1 \times 10^{-4})$ |
| $\chi_{b_0}(1^3P_0)$ | $0^{++}$ | 9.91134 | 9.898 | 9.847 | 9.859 | 9.8432 | 9.840 | $9.859 \pm (4.2 \times 10^{-4}) \pm (3.1 \times 10^{-4})$ |
| $h_b(1^1P_1)$ | $1^{+-}$ | 9.90400 | 9.918 | 9.882 | 9.900 | 9.87919 | 9.884 | $9.899 \pm (8 \times 10^{-4})$ |
| $\chi_{b_1}(2^3P_1)$ | $1^{++}$ | 10.2226 | 10.240 | 10.246 | 10.255 | 10.21695 | 10.229 | $10.255 \pm (2.2 \times 10^{-4}) \pm (5 \times 10^{-4})$ |
| $\chi_{b_2}(2^3P_2)$ | $2^{++}$ | 10.2178 | 10.245 | 10.261 | 10.268 | 10.22961 | 10.254 | $10.269 \pm (2.2 \times 10^{-4}) \pm (5 \times 10^{-4})$ |
| $\chi_{b_0}(2^3P_0)$ | $0^{++}$ | 10.2249 | 10.233 | 10.226 | 10.233 | 10.19625 | 10.202 | $10.233 \pm (4 \times 10^{-4}) \pm (5 \times 10^{-4})$ |
| $h_b(2^1P_1)^*$ | $1^{+-}$ | 10.2202 | 10.254 | 10.250 | 10.260 | 10.22153 | 10.237 | $10.260 \pm (1.2 \times 10^{-3})$ |
| $\Upsilon_2(1^3D_2)$ | $2^{--}$ | 10.1766 | 9.911 | 10.147 | 10.161 | 10.1126 | 10.096 | $10.164 \pm (1.4 \times 10^{-3})$ |
| $\Upsilon_3(1^3D_3)$ | $3^{--}$ | 10.1720 | 9.921 | 10.155 | 10.166 | 9.73855 | 9.849 | 10.172[28] |
| $\Upsilon_1(1^3D_1)$ | $1^{--}$ | 10.1796 | 9.900 | 10.138 | 10.154 | 9.72905 | 9.666 | 10.155[28] |
| $\eta_{b_2}(1^1D_2)$ | $2^{-+}$ | 10.1751 | 9.923 | 10.148 | 10.163 | 9.7355 | 9.767 | 10.165[28] |
| $\Upsilon_2(2^3D_2)$ | $2^{--}$ | 10.4038 | 10.244 | 10.449 | 10.443 | 9.94259 | 10.071 | |
| $\Upsilon_3(2^3D_3)$ | $3^{--}$ | 10.4005 | 10.253 | 10.455 | 10.449 | 9.94704 | 10.175 | |
| $\Upsilon_1(2^3D_1)$ | $1^{--}$ | 10.4060 | 10.235 | 10.441 | 10.435 | 9.93775 | 9.996 | |
| $\eta_{b_2}(2^1D_2)$ | $2^{-+}$ | 10.4027 | 10.258 | 10.450 | 10.445 | 9.94405 | 10.093 | |
| $\chi_{b_3}(1^3F_3)$ | $3^{++}$ | 10.3673 | 9.918 | 10.355 | 10.346 | 9.7361 | 9.754 | |
| $\chi_{b_4}(1^3F_4)$ | $4^{++}$ | 10.3640 | 9.932 | 10.358 | 10.349 | 9.74242 | 9.896 | |
| $\chi_{b_2}(1^3F_2)$ | $2^{++}$ | 10.3697 | 9.904 | 10.350 | 10.343 | 9.72948 | 9.642 | |
| $\eta_{b_3}(1^1F_3)$ | $3^{+-}$ | 10.3665 | 9.931 | 10.355 | 10.347 | 9.73759 | 9.778 | |
| $\chi_{b_3}(2^3F_3)$ | $3^{++}$ | 10.5457 | 10.251 | 10.619 | 10.614 | 9.94462 | 10.081 | |
| $\chi_{b_4}(2^3F_4)$ | $4^{++}$ | 10.5431 | 10.263 | 10.622 | 10.617 | 9.9508 | 10.219 | |
| $\chi_{b_2}(2^3F_2)$ | $2^{++}$ | 10.5476 | 10.239 | 10.615 | 10.610 | 9.93815 | 9.971 | |
| $\eta_{b_3}(2^1F_3)$ | $3^{+-}$ | 10.5450 | 10.265 | 10.619 | 10.615 | 9.94608 | 10.104 | |
| $\Upsilon_4(1^3G_4)$ | $4^{--}$ | 10.5155 | 9.929 | 10.531 | 10.512 | | | |
| $\Upsilon_5(1^3G_5)$ | $5^{--}$ | 10.5129 | 9.946 | 10.532 | 10.514 | | | |
| $\Upsilon_3(1^3G_3)$ | $3^{--}$ | 10.5175 | 9.911 | 10.529 | 10.511 | | | |
| $\eta_{b_4}(1^1G_4)$ | $4^{-+}$ | 10.5150 | 9.941 | 10.530 | 10.513 | | | |
| $APAD\%$ | | 0.25 | 1.56 | 0.13 | 0.011 | 1.08 | 1.06 | |

**Table 3**. Approximate mass spectrum of charmonium meson in $GeV$.

| State | | Present | Ref.[15] | Ref. [27] | Ref. [25] | Ref. [24] | Ref. [22] | Ref. [26] | Expt. [21] |
|---|---|---|---|---|---|---|---|---|---|
| $n^{2s+1}L_j$ | $J^{PC}$ | | | | | | | | |
| $J/\psi\,(1^3S_1)$ | $1^{--}$ | 3.0720 | 3.520 | 3.094 | 3.0413 | 3.096 | 3.126 | 3.0851 | $3.097 \pm (6 \times 10^{-6})$ |
| $\eta_c(1^1S_0)$ | $0^{-+}$ | 2.9861 | 3.293 | 2.989 | 3.1404 | 2.981 | 3.033 | 2.9904 | $2.984 \pm 0.0004$ |
| $\psi(2^3S_1)$ | $1^{--}$ | 3.5985 | 3.902 | 3.681 | 3.7017 | 3.685 | 3.701 | 3.6821 | $3.686 \pm (6 \times 10^{-5})$ |
| $\eta_c(2^1S_0)^*$ | $0^{-+}$ | 3.5877 | 3.638 | 3.602 | 3.6610 | 3.635 | 3.666 | 3.6465 | $3.638 \pm (1.1 \times 10^{-3})$ |
| $\chi_{c_1}(1^3P_1)^*$ | $1^{++}$ | 3.5556 | 3.511 | 3.468 | 3.5036 | 3.511 | 3.487 | 3.5004 | $3.511 \pm (5 \times 10^{-5})$ |
| $\chi_{c_2}(1^3P_2)$ | $2^{++}$ | 3.5388 | 3.545 | 3.480 | 3.4888 | 3.555 | 3.522 | 3.5514 | $3.556 \pm 7 \times 10^{-5}$ |
| $\chi_{c_0}(1^3P_0)$ | $0^{++}$ | 3.5619 | 3.466 | 3.428 | 3.4137 | 3.413 | 3.407 | 3.3519 | $3.415 \pm 3 \times 10^{-4}$ |
| $h_c(1^1P_1)$ | $1^{+-}$ | 3.5469 | 3.298 | 3.470 | 3.5180 | 3.525 | 3.502 | 3.5146 | $3.525 \pm (1.1 \times 10^{-4})$ |
| $\chi_{c_1}(2^3P_1)$ | $1^{++}$ | 3.8407 | 3.895 | 3.938 | 3.8072 | 3.906 | 3.786 | 3.9335 | $3.872 \pm (6 \times 10^{-5})$ |
| $\chi_{c_2}(2^3P_2)^*$ | $2^{++}$ | 3.8295 | 3.923 | 3.955 | 3.9151 | 3.949 | 3.905 | 3.9798 | $3.923 \pm 1 \times 10^{-3}$ |
| $\chi_{c_0}(2^3P_0)$ | $0^{++}$ | 3.8455 | 3.856 | 3.897 | 3.7646 | 3.870 | 3.899 | 3.8357 | $3.922 \pm 1.8 \times 10^{-3}$ |
| $h_c(2^1P_1)$ | $1^{+-}$ | 3.8350 | 3.642 | 3.943 | 3.8239 | 3.926 | 3.8210 | 3.9446 | |
| $\psi_2(1^3D_2)$ | $2^{--}$ | 3.7997 | 3.520 | 3.772 | 3.46047 | 3.795 | 3.3480 | 3.8077 | |
| $\psi_3(1^3D_3)$ | $3^{--}$ | 3.7884 | 3.575 | 3.755 | 3.51402 | 3.813 | 3.307 | 3.8146 | |
| $\psi_1(1^3D_1)$ | $1^{--}$ | 3.8069 | 3.461 | 3.775 | 3.40228 | 3.783 | 3.374 | 3.7853 | $3.774 \pm 4 \times 10^{-4}$ |
| $\eta_{c_2}(1^1D_2)$ | $2^{-+}$ | 3.7959 | 3.306 | 3.765 | 3.47795 | 3.807 | 3.3760 | 3.8073 | |
| $\psi_2(2^3D_2)$ | $2^{--}$ | 4.0008 | 3.902 | 4.188 | 3.81161 | 4.190 | 3.8010 | 4.1737 | $3.824 \pm (5 \times 10^{-4})$ |
| $\psi_3(2^3D_3)$ | $3^{--}$ | 3.9928 | 3.949 | 4.176 | 3.86300 | 4.220 | 3.797 | 4.1829 | |
| $\psi_1(2^3D_1)$ | $1^{--}$ | 4.0060 | 3.852 | 4.188 | 3.75606 | 4.150 | 3.800 | 4.1504 | |
| $\eta_{c_2}(2^1D_2)$ | $2^{-+}$ | 3.9982 | 3.649 | 4.182 | 3.82825 | 4.196 | 3.8360 | 4.1737 | |
| $\chi_{c_3}(1^3F_3)$ | $3^{++}$ | 3.9688 | 3.532 | 4.012 | 3.46746 | 4.068 | 3.375 | 4.0440 | |
| $\chi_{c_4}(1^3F_4)$ | $4^{++}$ | 3.9606 | 3.609 | 4.036 | 3.54185 | 4.093 | 3.315 | 4.0374 | |
| $\chi_{c_2}(1^3F_2)$ | $2^{++}$ | 3.9747 | 3.454 | 3.990 | 3.38961 | 4.041 | 3.403 | 4.0424 | |
| $\eta_{c_3}(1^1F_3)$ | $3^{+-}$ | 3.9668 | 3.319 | 4.017 | 3.48494 | 4.071 | 3.403 | 4.0411 | |
| $\chi_{c_3}(2^3F_3)$ | $3^{++}$ | 4.1253 | 3.913 | 4.396 | 3.81815 | 4.400 | 3.823 | 4.3744 | |
| $\chi_{c_4}(2^3F_4)$ | $4^{++}$ | 4.1191 | 3.978 | 4.415 | 3.88949 | 4.434 | 3.814 | 4.3711 | |
| $\chi_{c_2}(2^3F_2)$ | $2^{++}$ | 4.1299 | 3.846 | 4.378 | 3.74376 | 4.361 | 3.812 | 4.3699 | |
| $\eta_{c_3}(2^1F_3)$ | $3^{+-}$ | 4.1238 | 3.660 | 4.400 | 3.83479 | 4.406 | 3.858 | 4.3723 | |
| $\psi_{c_4}(1^3G_4)$ | $4^{--}$ | 4.0990 | 3.549 | | | 4.343 | | 4.2506 | |
| $\psi_{c_5}(1^3G_5)$ | $5^{--}$ | 4.0927 | 3.647 | | | 4.357 | | 4.2369 | |
| $\psi_{c_3}(1^3G_3)$ | $3^{--}$ | 4.1039 | 3.450 | | | 4.321 | | 4.2582 | |
| $\eta_{c_4}(1^1G_4)$ | $4^{-+}$ | 4.0978 | 3.336 | | | 4.345 | | 4.2471 | |
| $APAD\%$ | | 1.69 | 3.90 | 1.49 | 2.04 | 1.00 | 1.60 | 1.40 | |

**Table4**. Absolute deviation of quarkonia mass spectra in MeV.

| Bottomonium | | Charmonium | |
|---|---|---|---|
| State | Absolute deviation $\left\|M_{\exp} - M_{theor}\right\|$ | State | Absolute deviation $\left\|M_{\exp} - M_{theor}\right\|$ |
| $\boldsymbol{n^{2s+1}L_j}$ | | | |
| $\boldsymbol{\Upsilon(1^3S_1)}$ | 40.17 | $\boldsymbol{\psi_2(2^3D_2)}$ | 25.00 |
| $\boldsymbol{\eta_b(1^1S_0)}$ | 16.55 | $\boldsymbol{\psi_1(1^3D_1)}$ | 2.10 |
| $\boldsymbol{\Upsilon(2^3S_1)}$ | 62.49 | $\boldsymbol{J/\psi\ (1^3S_1)}$ | 87.50 |
| $\boldsymbol{\chi_{b_1}(1^3P_1)}$ | 14.78 | $\boldsymbol{\eta_c(1^1S_0)}$ | 50.30 |
| $\boldsymbol{\chi_{b_2}(1^3P_2)}$ | 11.78 | $\boldsymbol{\psi(2^3S_1)}$ | 44.60 |
| $\boldsymbol{\chi_{b_0}(1^3P_0)}$ | 52.34 | $\boldsymbol{\eta_c(2^1S_0)}$ | 17.20 |
| $\boldsymbol{h_b(1^1P_1)}$ | 5.00 | $\boldsymbol{\chi_{c_1}(1^3P_1)}$ | 146.90 |
| $\boldsymbol{\chi_{b_1}(2^3P_1)}$ | 32.40 | $\boldsymbol{\chi_{c_2}(1^3P_2)}$ | 21.90 |
| $\boldsymbol{\chi_{b_2}(2^3P_2)}$ | 51.20 | $\boldsymbol{\chi_{c_0}(1^3P_0)}$ | 31.30 |
| $\boldsymbol{\chi_{b_0}(2^3P_0)}$ | 8.10 | $\boldsymbol{h_c(1^1P_1)}$ | 93.50 |
| $\boldsymbol{h_b(2^1P_1)}$ | 39.80 | $\boldsymbol{\chi_{c_1}(2^3P_1)}$ | 76.50 |
| $\boldsymbol{\Upsilon_2(1^3D_2)}$ | 12.60 | $\boldsymbol{\chi_{c_2}(2^3P_2)}$ | 32.90 |
| $\boldsymbol{\Upsilon_3(1^3D_3)}$ | 0.00 | $\boldsymbol{\chi_{c_0}(2^3P_0)}$ | 176.80 |
| $\boldsymbol{\Upsilon_1(1^3D_1)}$ | 24.60 | | |
| $\boldsymbol{\eta_{b_2}(1^1D_2)}$ | 10.10 | | |

**Table 5**.Variations of energy eigenvalues ($E_{n0}$) in eV with potential parameter $B(eV)$ and radial quantum number. $\mu = 1(eV), \hbar = 1, \alpha_s = 2, l = 0, \lambda = 0.001(eV)$.

| | $n_r = 0$ | | | $n_r = 1$ | | |
|---|---|---|---|---|---|---|
| $B(eV)$ | Numerical [29] | Approx. | % error | Numerical [29] | Approx. | % error |
| 0.001 | -3.561574606 | -3.563214504 | 0.046044185 | -0.894899747 | -0.895104689 | 0.022901113 |
| 0.002 | -3.569233253 | -3.57087345 | 0.045953763 | -0.901116092 | -0.901320489 | 0.022682649 |
| 0.003 | -3.576892053 | -3.578532398 | 0.045859505 | -0.907332958 | -0.90753629 | 0.022409855 |
| 0.004 | -3.584551006 | -3.586191346 | 0.045761380 | -0.913550345 | -0.91375209 | 0.022083621 |
| 0.005 | -3.592210112 | -3.593850293 | 0.045659384 | -0.919967890 | -0.919768253 | 0.021700431 |
| 0.006 | -3.599869370 | -3.601509240 | 0.045553597 | -0.925986680 | -0.926183690 | 0.021275684 |
| 0.007 | -3.607528782 | -3.609168188 | 0.045444017 | -0.932205625 | -0.932399490 | 0.020796377 |
| 0.008 | -3.615188345 | -3.616827135 | 0.045330695 | -0.938425089 | 0.938615291 | 0.020268213 |
| 0.009 | -3.622848062 | -3.624486082 | 0.045213599 | -0.944645069 | -0.944831091 | 0.019692264 |
| 0.010 | -3.630507931 | -3.632145030 | 0.045092836 | -0.950865565 | -0.951046891 | 0.019069573 |

**Table 6.** Variations of energy eigenvalues ($E_{nl}$) in eV for fixed potential parameters with $l > 0$, $B = 0.001(eV), \mu = 1\text{eV}, \hbar = 1, \alpha_s = 2, \lambda = 0.001(eV)$.

| | $n_r = 0$ | | | $n_r = 1$ | | |
|---|---|---|---|---|---|---|
| $l$ | Numerical [29] | Approx. | % error | Numerical [29] | Approx. | % error |
| 1 | -0.895272060 | -0.895271356 | 0.0000786353 | -0.400539503 | -0.400538730 | 0.000192990 |
| 2 | -0.400672719 | -0.4006720635 | 0.000163600 | -0.227171016 | -0.2271698753 | 0.000502133 |
| 3 | -0.227278172 | -0.227277018 | 0.000507748 | -0.146755665 | -0.146753875 | 0.001219714 |
| 4 | -0.146844555 | -0.146842763 | 0.001220338 | -0.102952464 | -0.102953652 | 0.001153931 |
| 5 | -0.103031297 | -0.103029409 | 0.001832453 | -0.076227867 | -0.076457899 | 0.301768906 |

**Table 7**.Variations of energy eigenvalues ($E_{n0}$) in eV with potential parameter $\alpha_s$ and radial quantum number. $\mu = 1(eV), \hbar = 1, l = 0, \lambda = 0.001(eV), B = 0.001(eV)$.

| | $n_r = 0$ | | | $n_r = 1$ | | |
|---|---|---|---|---|---|---|
| $\alpha_s$ | Numerical [29] | Approx. | % error | Numerical [29] | Approx. | % error |
| 3 | -7.999882716 | -8.008064413 | 0.102272712 | -2.005597913 | -2.006621266 | 0.051024834 |
| 4 | -14.20509595 | -14.23057432 | 0.179360774 | -3.559277093 | -3.562464504 | 0.089552202 |
| 5 | -22.16952050 | -22.23079746 | 0.276401828 | -5.555019617 | -5.562687647 | 0.138037856 |
| 6 | -31.88360823 | -32.00875756 | 0.392519344 | -7.991647820 | -8.007314413 | 0.196037080 |
| 7 | -43.33614208 | -43.56446727 | 0.526870134 | -10.86776228 | -10.89635746 | 0.263119300 |

## 4. Conclusion

In this article, we derived an approximate energy equation for a Coulomb plus logarithmic potential using perturbation theory. The resulting energy spectra to first-order corrections, were applied to the mass spectra of bottomonium and charmonium mesons. This enabled the determination of the free potential parameters through fitting to experimental data of the Particle Data Group. The proposed potential, together with the modeled parameters, accounts for asymptotic freedom at short distances through one-gluon exchange interactions between quark–antiquark pairs, as well as quark confinement due to logarithmic function which grows with distance. For the bottomonium system, an absolute percentage average deviation (APAD) of 0.25% was obtained relative to the PDG mass values. This result represents an improvement over previous studies that reported APAD values of 1.56% [15], 1.06% [22], and 1.08% [25], where the Pekeris approximation and other phenomenological potentials were employed. The vector and pseudoscalar masses of charmonium were also calculated. The predicted masses showed good agreement with experimental data [21], yielding an APAD of 1.69%. These results improve upon earlier reports in Refs. [15, 24] and are comparable with other competing theoretical predictions [22, 21–27] for higher orbital quantum number ($l$). Also, the ground-state and first excited-state energy eigenvalues for arbitrary potential parameters in the absence of spin components agree with the numerical solutions calculated using the matrix Numerov method.

## ACKNOWLEDGEMENTS

E. P. Inyang gratefully acknowledges the 2025 TETFund Institutional Based Research (IBR) grant of the National Open University of Nigeria for providing support for this study.

## FUNDING INFORMATION

This research was funded by the 2025 TETFund Institutional Based Research (IBR) grant of the National Open University of Nigeria with grant number NOUN/TETFIBR25/FACSCIENCES/063.

## CREDIT AUTHORSHIP CONTRIBUTION STATEMENT

E.O. Conceptualization, Data curation, Methodology, Writing – original draft, Writing – review & editing. E.P.I. Writing – review & editing, Visualization. K. A. Formal analysis & editing E.M.K. Conceptualization, Formal analysis, Writing – original draft, Writing – review & editing. All authors reviewed the manuscript.

## DATA AVAILABILITY STATEMENT

This manuscript has no associated data. The data used in this work are gotten from the derived equations within the article and the cited references.

**CONSENT TO PUBLISH DECLARATION**: Not applicable

**CONSENT TO PARTICIPATE DECLARATION**: Not applicable

**ETHICS DECLARATION**: Not applicable

**CONFLICT OF INTEREST:** The authors declare that they have no known competing financial interests or personal relationships that could have appeared to influence the work reported in this paper.

## APENDIX

The approximate mass spectra equation is given as

$$M_i^{theo.}\left(B,\alpha_s,\lambda,\sigma\right)=m_q+m_{\bar{q}}+E_{n_rl}^{(0)}+E_{n_rl}^{(1)} \qquad (A_1)$$

The minimization requires that

$$\sum_{i=1}^{K}\left(M_i^{\exp.}-M_i^{theo.}\left(B,\alpha_s,\lambda,\sigma\right)\right)^2=0 \qquad (A_2)$$

where $K$ is the number of experimental data points. Using the NLPSolve function in Maple program, we have obtained the potential parameters from eq. (**$A_2$**). We give example for the charmonium spectrum below:

Vars := [sigma, lambda, alpha, B];

```
with(Optimization);

L := 1/2*(j*(j+1)-l*(l+1)-s*(s+1));
T[2] := -(1/6)*l/(2*l+3);
T[1] := 1/6;
T[0] := (1/6)*(-1-l)/(2*l-1);
A[0] := B*L/(2*M*m);
A[1] := 2*alpha*(2*T[0]+L)/(M m);
a := sqrt((l+1/2)^2-2*mu*A[0]/h^2);
b := sqrt(2*mu/h^2*(8*alpha^2*mu/(9*h^2*(n+1/2+sqrt((l+1/2)^2-2*mu*A[0]/h^2))^2)));
Lambda := `&x`(16*alpha*Pi*(sigma/sqrt(Pi))^3/(9*M*m), s*(s+1)-3/2);
mu := M*m/(M+m);
M[1]:=M+m-8*alpha^2*mu/(9*h^2*(n+1/2+sqrt((l+1/2)^2-
2*mu*A[0]/h^2))^2)+1*(B*(ln(lambda/(2*b))+Psi(0,
2*a+2))+`&x`(Lambda*b*(2*b/sigma)^(2*a+1)/(sigma^2*GAMMA(2*a+2)),
sigma*GAMMA(a+1)*Hypergeom([a+1], [1/2], b^2/sigma^2)-2*b*GAMMA(a+3/2)*Hypergeom([a+3/2],
[3/2], b^2/sigma^2))+4*b^3*A[1]/(4*a^3-a));

M[2]:=M+m-8*alpha^2*mu/(9*h^2*(n+1/2+sqrt((l+1/2)^2-
2*mu*A[0]/h^2))^2)+1*(B*(3/(2*a+3)+ln(lambda/(2*b))+Psi(0,
2*a+2))+`&x`(Lambda*b*(2*b/sigma)^(2*a+1)/(sigma^4*(2*a+3)*GAMMA(2*a+2)),
sigma*GAMMA(a+1)*((2*a*sigma+sigma)^2*Hypergeom([a+1],[1/2],
b^2/sigma^2)+4*b^2*(a+1)*(Hypergeom([a+2], [1/2], b^2/sigma^2)+(2*(2*a+1))*Hypergeom([a+2], [3/2],
b^2/sigma^2)))-2*b*GAMMA(a+3/2)*(2*sigma^2*(2*a+1)*Hypergeom([a+3/2],[1/2],
b^2/sigma^2)+(2*a*sigma+sigma)^2*Hypergeom([a+3/2],[3/2],
b^2/sigma^2)+2*b^2*(2*a+3)*Hypergeom([a+5/2], [3/2], b^2/sigma^2)))+4*b^3*A[1]/(4*a^3-a));

t1:=-3.097+3.04-.6755555556*alpha^2+B*(ln(.4934210528*lambda/sqrt(alpha^2))+3/2-
gamma)+.8006584365*alpha^3*sqrt(alpha^2)*((1/2)*sigma*sqrt(Pi)*Hypergeom([3/2],[1/2],
1.026844444*alpha^2/sigma^2)-2.026666667*sqrt(alpha^2)*Hypergeom([2],[3/2],
1.026844444*alpha^2/sigma^2))/(sqrt(Pi)*sigma);

    t2 := -2.984+3.04-.6755555556*alpha^2+B*(ln(.4934210528*lambda/sqrt(alpha^2))+3/2-gamma)-
    2.401975308*alpha^3*sqrt(alpha^2)*((1/2)*sigma*sqrt(Pi)*Hypergeom([3/2], [1/2],
    1.026844444*alpha^2/sigma^2)-2.026666667*sqrt(alpha^2)*Hypergeom([2], [3/2],
    1.026844444*alpha^2/sigma^2))/(sqrt(Pi)*sigma);

    t3 := -3.686-.1688888889*alpha^2+B*(9/4+ln(.9868421051*lambda/sqrt(alpha^2))-
    gamma)+0.2502057612e-
    1*alpha^3*sqrt(alpha^2)*((1/2)*sigma*sqrt(Pi)*(4*sigma^2*Hypergeom([3/2], [1/2],
    .2567111111*alpha^2/sigma^2)+1.540266667*alpha^2*(Hypergeom([5/2], [1/2],
    .2567111111*alpha^2/sigma^2)+4*Hypergeom([5/2], [3/2], .2567111111*alpha^2/sigma^2)))-
    1.013333333*sqrt(alpha^2)*(4*sigma^2*Hypergeom([2], [1/2],
```

```
.2567111111*alpha^2/sigma^2)+4*sigma^2*Hypergeom([2], [3/2],
.2567111111*alpha^2/sigma^2)+2.053688889*alpha^2*Hypergeom([3], [3/2],
.2567111111*alpha^2/sigma^2)))/(sqrt(Pi)*sigma^3)+3.04;

t4 := -3.638-.1688888889*alpha^2+B*(9/4+ln(.9868421051*lambda/sqrt(alpha^2))-gamma)-
0.7506172836e-1*alpha^3*sqrt(alpha^2)*((1/2)*sigma*sqrt(Pi)*(4*sigma^2*Hypergeom([3/2],
[1/2], .2567111111*alpha^2/sigma^2)+1.540266667*alpha^2*(Hypergeom([5/2], [1/2],
.2567111111*alpha^2/sigma^2)+4*Hypergeom([5/2], [3/2], .2567111111*alpha^2/sigma^2)))-
1.013333333*sqrt(alpha^2)*(4*sigma^2*Hypergeom([2], [1/2],
.2567111111*alpha^2/sigma^2)+4*sigma^2*Hypergeom([2], [3/2],
.2567111111*alpha^2/sigma^2)+2.053688889*alpha^2*Hypergeom([3], [3/2],
.2567111111*alpha^2/sigma^2)))/(sqrt(Pi)*sigma^3)+3.04;

t5 := -3.511+3.04-
0.6755555556*alpha^2/(1/2+sqrt(9/4+.3289473684*B))^2+B*(ln(.4934210528*lambda/sqrt(alpha
^2/(1/2+sqrt(9/4+.3289473684*B))^2))+Psi(2*sqrt(9/4+.3289473684*B)+2))+.3898635477*alpha*
sigma*sqrt(alpha^2/(1/2+sqrt(9/4+.3289473684*B))^2)*(2.026666667*sqrt(alpha^2/(1/2+sqrt(9/4+
.3289473684*B))^2)/sigma)^(2*sqrt(9/4+.3289473684*B)+1)*(sigma*GAMMA(sqrt(9/4+.328947
3684*B)+1)*Hypergeom([sqrt(9/4+.3289473684*B)+1], [1/2],
1.026844444*alpha^2/((1/2+sqrt(9/4+.3289473684*B))^2*sigma^2))-
2.026666667*sqrt(alpha^2/(1/2+sqrt(9/4+.3289473684*B))^2)*GAMMA(sqrt(9/4+.3289473684*B
)+3/2)*Hypergeom([sqrt(9/4+.3289473684*B)+3/2], [3/2],
1.026844444*alpha^2/((1/2+sqrt(9/4+.3289473684*B))^2*sigma^2)))/(sqrt(Pi)*GAMMA(2*sqrt(9
/4+.3289473684*B)+2))-
2.401975309*(alpha^2/(1/2+sqrt(9/4+.3289473684*B))^2)^(3/2)*alpha/(4*(9/4+.3289473684*B)^(
3/2)-sqrt(9/4+.3289473684*B));

t6 := -3.556+3.04-.6755555556*alpha^2/(1/2+sqrt(9/4-
.3289473684*B))^2+B*(ln(.4934210528*lambda/sqrt(alpha^2/(1/2+sqrt(9/4-
.3289473684*B))^2))+Psi(2*sqrt(9/4-
.3289473684*B)+2))+.3898635477*alpha*sigma*sqrt(alpha^2/(1/2+sqrt(9/4-
.3289473684*B))^2)*(2.026666667*sqrt(alpha^2/(1/2+sqrt(9/4-
.3289473684*B))^2)/sigma)^(2*sqrt(9/4-.3289473684*B)+1)*(sigma*GAMMA(sqrt(9/4-
.3289473684*B)+1)*Hypergeom([sqrt(9/4-.3289473684*B)+1], [1/2],
1.026844444*alpha^2/((1/2+sqrt(9/4-.3289473684*B))^2*sigma^2))-
2.026666667*sqrt(alpha^2/(1/2+sqrt(9/4-.3289473684*B))^2)*GAMMA(sqrt(9/4-
.3289473684*B)+3/2)*Hypergeom([sqrt(9/4-.3289473684*B)+3/2], [3/2],
1.026844444*alpha^2/((1/2+sqrt(9/4-
.3289473684*B))^2*sigma^2)))/(sqrt(Pi)*GAMMA(2*sqrt(9/4-
.3289473684*B)+2))+3.362765431*(alpha^2/(1/2+sqrt(9/4-
.3289473684*B))^2)^(3/2)*alpha/(4*(9/4-.3289473684*B)^(3/2)-sqrt(9/4-.3289473684*B));

t7 := -3.415+3.04-
.6755555556*alpha^2/(1/2+sqrt(9/4+.6578947368*B))^2+B*(ln(.4934210528*lambda/sqrt(alpha^
2/(1/2+sqrt(9/4+.6578947368*B))^2))+Psi(2*sqrt(9/4+.6578947368*B)+2))+.3898635477*alpha*s
igma*sqrt(alpha^2/(1/2+sqrt(9/4+.6578947368*B))^2)*(2.026666667*sqrt(alpha^2/(1/2+sqrt(9/4+.
6578947368*B))^2)/sigma)^(2*sqrt(9/4+.6578947368*B)+1)*(sigma*GAMMA(sqrt(9/4+.657894
```

```
7368*B)+1)*Hypergeom([sqrt(9/4+.6578947368*B)+1], [1/2],
1.026844444*alpha^2/((1/2+sqrt(9/4+.6578947368*B))^2*sigma^2))-
2.026666667*sqrt(alpha^2/(1/2+sqrt(9/4+.6578947368*B))^2)*GAMMA(sqrt(9/4+.6578947368*B
)+3/2)*Hypergeom([sqrt(9/4+.6578947368*B)+3/2], [3/2],
1.026844444*alpha^2/((1/2+sqrt(9/4+.6578947368*B))^2*sigma^2)))/(sqrt(Pi)*GAMMA(2*sqrt(9
/4+.6578947368*B)+2))-
9.607901237*(alpha^2/(1/2+sqrt(9/4+.6578947368*B))^2)^(3/2)*alpha/(4*(9/4+.6578947368*B)^(
3/2)-sqrt(9/4+.6578947368*B));

t8 := -3.525+3.04-.1688888889*alpha^2+B*(ln(.9868421051*lambda/sqrt(alpha^2))+25/12-
gamma)-0.2569223957e-1*alpha^5*sqrt(alpha^2)*((3/4)*sigma*sqrt(Pi)*Hypergeom([5/2], [1/2],
.2567111111*alpha^2/sigma^2)-2.026666667*sqrt(alpha^2)*Hypergeom([3], [3/2],
.2567111111*alpha^2/sigma^2))/(sqrt(Pi)*sigma^3);

t9 := -3.872-
.6755555556*alpha^2/(sqrt(9/4+.3289473684*B)+3/2)^2+B*(3/(2*sqrt(9/4+.3289473684*B)+3)+l
n(.4934210528*lambda/sqrt(alpha^2/(sqrt(9/4+.3289473684*B)+3/2)^2))+Psi(2*sqrt(9/4+.328947
3684*B)+2))+.3898635477*alpha*sqrt(alpha^2/(sqrt(9/4+.3289473684*B)+3/2)^2)*(2.026666667
*sqrt(alpha^2/(sqrt(9/4+.3289473684*B)+3/2)^2)/sigma)^(2*sqrt(9/4+.3289473684*B)+1)*(sigma
*GAMMA(sqrt(9/4+.3289473684*B)+1)*((2*sqrt(9/4+.3289473684*B)*sigma+sigma)^2*Hyperg
eom([sqrt(9/4+.3289473684*B)+1], [1/2],
1.026844444*alpha^2/((sqrt(9/4+.3289473684*B)+3/2)^2*sigma^2))+4.107377778*alpha^2*(sqrt(
9/4+.3289473684*B)+1)*(Hypergeom([sqrt(9/4+.3289473684*B)+2], [1/2],
1.026844444*alpha^2/((sqrt(9/4+.3289473684*B)+3/2)^2*sigma^2))+(2*(2*sqrt(9/4+.3289473684
*B)+1))*Hypergeom([sqrt(9/4+.3289473684*B)+2], [3/2],
1.026844444*alpha^2/((sqrt(9/4+.3289473684*B)+3/2)^2*sigma^2)))/(sqrt(9/4+.3289473684*B)+
3/2)^2)-
2.026666667*sqrt(alpha^2/(sqrt(9/4+.3289473684*B)+3/2)^2)*GAMMA(sqrt(9/4+.3289473684*B
)+3/2)*(2*sigma^2*(2*sqrt(9/4+.3289473684*B)+1)*Hypergeom([sqrt(9/4+.3289473684*B)+3/2],
[1/2],
1.026844444*alpha^2/((sqrt(9/4+.3289473684*B)+3/2)^2*sigma^2))+(2*sqrt(9/4+.3289473684*B
)*sigma+sigma)^2*Hypergeom([sqrt(9/4+.3289473684*B)+3/2], [3/2],
1.026844444*alpha^2/((sqrt(9/4+.3289473684*B)+3/2)^2*sigma^2))+2.053688889*alpha^2*(2*sq
rt(9/4+.3289473684*B)+3)*Hypergeom([sqrt(9/4+.3289473684*B)+5/2], [3/2],
1.026844444*alpha^2/((sqrt(9/4+.3289473684*B)+3/2)^2*sigma^2))/(sqrt(9/4+.3289473684*B)+3
/2)^2))/(sqrt(Pi)*sigma*(2*sqrt(9/4+.3289473684*B)+3)*GAMMA(2*sqrt(9/4+.3289473684*B)+
2))-
2.401975309*(alpha^2/(sqrt(9/4+.3289473684*B)+3/2)^2)^(3/2)*alpha/(4*(9/4+.3289473684*B)^(
3/2)-sqrt(9/4+.3289473684*B))+3.04;

t10 := -3.923-.6755555556*alpha^2/(sqrt(9/4-.3289473684*B)+3/2)^2+B*(3/(2*sqrt(9/4-
.3289473684*B)+3)+ln(.4934210528*lambda/sqrt(alpha^2/(sqrt(9/4-
.3289473684*B)+3/2)^2))+Psi(2*sqrt(9/4-
.3289473684*B)+2))+.3898635477*alpha*sqrt(alpha^2/(sqrt(9/4-
.3289473684*B)+3/2)^2)*(2.026666667*sqrt(alpha^2/(sqrt(9/4-
.3289473684*B)+3/2)^2)/sigma)^(2*sqrt(9/4-.3289473684*B)+1)*(sigma*GAMMA(sqrt(9/4-
```

```
.3289473684*B)+1)*((2*sqrt(9/4-.3289473684*B)*sigma+sigma)^2*Hypergeom([sqrt(9/4-
.3289473684*B)+1], [1/2], 1.026844444*alpha^2/((sqrt(9/4-
.3289473684*B)+3/2)^2*sigma^2))+4.107377778*alpha^2*(sqrt(9/4-
.3289473684*B)+1)*(Hypergeom([sqrt(9/4-.3289473684*B)+2], [1/2],
1.026844444*alpha^2/((sqrt(9/4-.3289473684*B)+3/2)^2*sigma^2))+(2*(2*sqrt(9/4-
.3289473684*B)+1))*Hypergeom([sqrt(9/4-.3289473684*B)+2], [3/2],
1.026844444*alpha^2/((sqrt(9/4-.3289473684*B)+3/2)^2*sigma^2)))/(sqrt(9/4-
.3289473684*B)+3/2)^2)-2.026666667*sqrt(alpha^2/(sqrt(9/4-
.3289473684*B)+3/2)^2)*GAMMA(sqrt(9/4-.3289473684*B)+3/2)*(2*sigma^2*(2*sqrt(9/4-
.3289473684*B)+1)*Hypergeom([sqrt(9/4-.3289473684*B)+3/2], [1/2],
1.026844444*alpha^2/((sqrt(9/4-.3289473684*B)+3/2)^2*sigma^2))+(2*sqrt(9/4-
.3289473684*B)*sigma+sigma)^2*Hypergeom([sqrt(9/4-.3289473684*B)+3/2], [3/2],
1.026844444*alpha^2/((sqrt(9/4-
.3289473684*B)+3/2)^2*sigma^2))+2.053688889*alpha^2*(2*sqrt(9/4-
.3289473684*B)+3)*Hypergeom([sqrt(9/4-.3289473684*B)+5/2], [3/2],
1.026844444*alpha^2/((sqrt(9/4-.3289473684*B)+3/2)^2*sigma^2))/(sqrt(9/4-
.3289473684*B)+3/2)^2))/(sqrt(Pi)*sigma*(2*sqrt(9/4-.3289473684*B)+3)*GAMMA(2*sqrt(9/4-
.3289473684*B)+2))+3.362765431*(alpha^2/(sqrt(9/4-
.3289473684*B)+3/2)^2)^(3/2)*alpha/(4*(9/4-.3289473684*B)^(3/2)-sqrt(9/4-
.3289473684*B))+3.04;

t11 := -3.922-
.6755555556*alpha^2/(sqrt(9/4+.6578947368*B)+3/2)^2+B*(3/(2*sqrt(9/4+.6578947368*B)+3)+l
n(.4934210528*lambda/sqrt(alpha^2/(sqrt(9/4+.6578947368*B)+3/2)^2))+Psi(2*sqrt(9/4+.657894
7368*B)+2))+.3898635477*alpha*sqrt(alpha^2/(sqrt(9/4+.6578947368*B)+3/2)^2)*(2.026666667
*sqrt(alpha^2/(sqrt(9/4+.6578947368*B)+3/2)^2)/sigma)^(2*sqrt(9/4+.6578947368*B)+1)*(sigma
*GAMMA(sqrt(9/4+.6578947368*B)+1)*((2*sqrt(9/4+.6578947368*B)*sigma+sigma)^2*Hyperg
eom([sqrt(9/4+.6578947368*B)+1], [1/2],
1.026844444*alpha^2/((sqrt(9/4+.6578947368*B)+3/2)^2*sigma^2))+4.107377778*alpha^2*(sqrt(
9/4+.6578947368*B)+1)*(Hypergeom([sqrt(9/4+.6578947368*B)+2], [1/2],
1.026844444*alpha^2/((sqrt(9/4+.6578947368*B)+3/2)^2*sigma^2))+(2*(2*sqrt(9/4+.6578947368
*B)+1))*Hypergeom([sqrt(9/4+.6578947368*B)+2], [3/2],
1.026844444*alpha^2/((sqrt(9/4+.6578947368*B)+3/2)^2*sigma^2)))/(sqrt(9/4+.6578947368*B)+
3/2)^2)-
2.026666667*sqrt(alpha^2/(sqrt(9/4+.6578947368*B)+3/2)^2)*GAMMA(sqrt(9/4+.6578947368*B
)+3/2)*(2*sigma^2*(2*sqrt(9/4+.6578947368*B)+1)*Hypergeom([sqrt(9/4+.6578947368*B)+3/2],
[1/2],
1.026844444*alpha^2/((sqrt(9/4+.6578947368*B)+3/2)^2*sigma^2))+(2*sqrt(9/4+.6578947368*B
)*sigma+sigma)^2*Hypergeom([sqrt(9/4+.6578947368*B)+3/2], [3/2],
1.026844444*alpha^2/((sqrt(9/4+.6578947368*B)+3/2)^2*sigma^2))+2.053688889*alpha^2*(2*sq
rt(9/4+.6578947368*B)+3)*Hypergeom([sqrt(9/4+.6578947368*B)+5/2], [3/2],
1.026844444*alpha^2/((sqrt(9/4+.6578947368*B)+3/2)^2*sigma^2))/(sqrt(9/4+.6578947368*B)+3
/2)^2))/(sqrt(Pi)*sigma*(2*sqrt(9/4+.6578947368*B)+3)*GAMMA(2*sqrt(9/4+.6578947368*B)+
2))-
9.607901237*(alpha^2/(sqrt(9/4+.6578947368*B)+3/2)^2)^(3/2)*alpha/(4*(9/4+.6578947368*B)^(
3/2)-sqrt(9/4+.6578947368*B))+3.04;
```

```
t12 := -3.774+3.04-
.6755555556*alpha^2/(1/2+sqrt(25/4+.9868421052*B))^2+B*(ln(.4934210528*lambda/sqrt(alpha
^2/(1/2+sqrt(25/4+.9868421052*B))^2))+Psi(2*sqrt(25/4+.9868421052*B)+2))+.3898635477*alph
a*sigma*sqrt(alpha^2/(1/2+sqrt(25/4+.9868421052*B))^2)*(2.026666667*sqrt(alpha^2/(1/2+sqrt(2
5/4+.9868421052*B))^2)/sigma)^(2*sqrt(25/4+.9868421052*B)+1)*(sigma*GAMMA(sqrt(25/4+.
9868421052*B)+1)*Hypergeom([sqrt(25/4+.9868421052*B)+1], [1/2],
1.026844444*alpha^2/((1/2+sqrt(25/4+.9868421052*B))^2*sigma^2))-
2.026666667*sqrt(alpha^2/(1/2+sqrt(25/4+.9868421052*B))^2)*GAMMA(sqrt(25/4+.9868421052
*B)+3/2)*Hypergeom([sqrt(25/4+.9868421052*B)+3/2], [3/2],
1.026844444*alpha^2/((1/2+sqrt(25/4+.9868421052*B))^2*sigma^2)))/(sqrt(Pi)*GAMMA(2*sqrt(
25/4+.9868421052*B)+2))-
12.00987654*(alpha^2/(1/2+sqrt(25/4+.9868421052*B))^2)^(3/2)*alpha/(4*(25/4+.9868421052*B
)^(3/2)-sqrt(25/4+.9868421052*B));

t13 := -3.824-
.6755555556*alpha^2/(sqrt(25/4+.3289473684*B)+3/2)^2+B*(3/(2*sqrt(25/4+.3289473684*B)+3)
+ln(.4934210528*lambda/sqrt(alpha^2/(sqrt(25/4+.3289473684*B)+3/2)^2))+Psi(2*sqrt(25/4+.328
9473684*B)+2))+.3898635477*alpha*sqrt(alpha^2/(sqrt(25/4+.3289473684*B)+3/2)^2)*(2.02666
6667*sqrt(alpha^2/(sqrt(25/4+.3289473684*B)+3/2)^2)/sigma)^(2*sqrt(25/4+.3289473684*B)+1)*
(sigma*GAMMA(sqrt(25/4+.3289473684*B)+1)*((2*sqrt(25/4+.3289473684*B)*sigma+sigma)^2
*Hypergeom([sqrt(25/4+.3289473684*B)+1], [1/2],
1.026844444*alpha^2/((sqrt(25/4+.3289473684*B)+3/2)^2*sigma^2))+4.107377778*alpha^2*(sqr
t(25/4+.3289473684*B)+1)*(Hypergeom([sqrt(25/4+.3289473684*B)+2], [1/2],
1.026844444*alpha^2/((sqrt(25/4+.3289473684*B)+3/2)^2*sigma^2))+(2*(2*sqrt(25/4+.32894736
84*B)+1))*Hypergeom([sqrt(25/4+.3289473684*B)+2], [3/2],
1.026844444*alpha^2/((sqrt(25/4+.3289473684*B)+3/2)^2*sigma^2)))/(sqrt(25/4+.3289473684*B
)+3/2)^2)-
2.026666667*sqrt(alpha^2/(sqrt(25/4+.3289473684*B)+3/2)^2)*GAMMA(sqrt(25/4+.3289473684
*B)+3/2)*(2*sigma^2*(2*sqrt(25/4+.3289473684*B)+1)*Hypergeom([sqrt(25/4+.3289473684*B)
+3/2], [1/2],
1.026844444*alpha^2/((sqrt(25/4+.3289473684*B)+3/2)^2*sigma^2))+(2*sqrt(25/4+.3289473684
*B)*sigma+sigma)^2*Hypergeom([sqrt(25/4+.3289473684*B)+3/2], [3/2],
1.026844444*alpha^2/((sqrt(25/4+.3289473684*B)+3/2)^2*sigma^2))+2.053688889*alpha^2*(2*s
qrt(25/4+.3289473684*B)+3)*Hypergeom([sqrt(25/4+.3289473684*B)+5/2], [3/2],
1.026844444*alpha^2/((sqrt(25/4+.3289473684*B)+3/2)^2*sigma^2))/(sqrt(25/4+.3289473684*B)
+3/2)^2))/(sqrt(Pi)*sigma*(2*sqrt(25/4+.3289473684*B)+3)*GAMMA(2*sqrt(25/4+.3289473684
*B)+2))-
2.401975309*(alpha^2/(sqrt(25/4+.3289473684*B)+3/2)^2)^(3/2)*alpha/(4*(25/4+.3289473684*B
)^(3/2)-sqrt(25/4+.3289473684*B))+3.04;

cc := t1^2+t2^2+t3^2+t4^2+t5^2+t6^2+t7^2+t8^2+t9^2;
sol := NLPSolve(cc, {alpha >= 0, lambda >= 0, sigma >= 0}, initialpoint = {B = .14, alpha = .8,
lambda = 1, sigma = 0.23});
```